# A Kolmogorov-Smirnov test for the molecular clock on Bayesian ensembles of phylogenies

Fernando Antoneli[1,2,¶], Fernando M. Passos[1,3,¶], Luciano R. Lopes[1,2] and Marcelo R. S. Briones[1,2,*]

[1]Laboratório de Genômica Evolutiva e Biocomplexidade, Escola Paulista de Medicina, Universidade Federal de São Paulo, Rua Pedro de Toledo, 669, Ed. Pesquisas 2, UNIFESP, São Paulo, SP, Brazil

[2]Departamento de Informática em Saúde, Escola Paulista de Medicina, Universidade Federal de São Paulo, Rua Botucatu 740, Ed. José Leal Prado, São Paulo, SP, Brazil

[3]Departamento de Análises Clínicas e Toxicológicas, Faculdade de Ciências Farmacêuticas, Universidade de São Paulo, Av. Professor Lineu Prestes 580, São Paulo, SP, Brazil

*Corresponding author
E-mail: marcelo.briones@unifesp.br
¶These authors contributed equally.

**Keywords:** Molecular clock, Molecular phylogeny, Phylogenetic tree, Evolution, Non-parametric goodness-of-fit test, Bayesian inference, Poisson process.

## Abstract

Divergence date estimates are central to understand evolutionary processes and depend, in the case of molecular phylogenies, on tests of molecular clocks. Here we propose two non-parametric tests of strict and relaxed molecular clocks built upon a framework that uses the empirical cumulative distribution (ECD) of branch lengths obtained from an ensemble of Bayesian trees and well known non-parametric (one-sample and two-sample) Kolmogorov-Smirnov (KS) goodness-of-fit test. In the strict clock case, the method consists in using the one-sample Kolmogorov-Smirnov (KS) test to directly test if the phylogeny is clock-like, in other words, if it follows a Poisson law. The ECD is computed from the discretized branch lengths and the parameter $\lambda$ of the expected Poisson distribution is calculated as the average branch length over the ensemble of trees. To compensate for the auto-correlation in the ensemble of trees and pseudo-replication we take advantage of thinning and effective sample size, two features provided by Bayesian inference MCMC samplers. Finally, it is observed that tree topologies with very long or very short branches lead to Poisson mixtures and in this case we propose the use of the two-sample KS test with samples from two continuous branch length distributions, one obtained from an ensemble of clock-constrained trees and the other from an ensemble of unconstrained trees. Moreover, in this second form the test can also be applied to test for relaxed clock models. The use of a statistically equivalent ensemble of phylogenies to obtain the branch lengths ECD, instead of one consensus tree, yields considerable reduction of the effects of small sample size and provides a gain of power.

## Introduction

The molecular clock hypothesis postulates that for a given informational macro-molecule (DNA or protein sequence) the evolutionary rate is approximately constant over time in all evolutionary lines of descent. This implies that if genetic divergence accumulates in a stochastic clock-like manner, that is, approximately constant number of mutations accumulated per time interval, then, time scales could be determined for evolutionary events, with calibration using fossil evidence. Moreover, the evolutionary rate variation between lineages could shed light on mechanisms of molecular evolution.

If the substitution rate is constant between lineages, then evolutionary distances are such that all external nodes of a phylogenetic tree should be of the same size starting from the root. In [1–3] authors suggest that the substitution process is approximately Poisson, meaning that the average number of substitutions, and its variance, in different lineages during the same time interval should be equal.

When the neutral theory of molecular evolution was proposed [4,5], the observed clock-like behavior of molecular evolution was advocated as strong evidence supporting the theory [4–6]. However, the reliability of the clock and its implications for the mechanism of molecular evolution were a focus of immediate controversy, entwined in the "neutralist–selectionist" controversy. The debate surrounding the neutral theory has generated a rich body of population genetics theory and analytical tools. For instance, in the strict neutral model the dynamics depends on the neutral mutation rate alone, however one may expect most sites in a functional protein to be constrained during most of the evolutionary time. This observation motivated the introduction of doubly stochastic Poisson process, or Cox process, as a model for the substitution process, implying that positive selection, if it occurs, is in episodic fashion and should affect only a few sites [7,8]. More recently, these ideas motivated the introduction of relaxed molecular clock models and advanced their use for inferring dates of divergence events.

Despite the great impact of the molecular clock in evolutionary biology, as comparative molecular data have been accumulating over the past decades, no prediction has been proven satisfactory; the dispersion index (the ratio of the variance to the mean value of the number of substitutions) is generally greater than 1, suggesting that the substitution process is over-dispersed. Furthermore, it was observed that the substitution rate usually display variation along lineages, a fact that has created great controversy on its use in divergence date estimates [9].

In order to address these issues, several statistical tests have been developed to examine whether rates of molecular evolution vary significantly among phylogenetic lineages. Fitch proposed a simple test for statistically examining whether the observed difference in evolutionary rates between two sequences is significantly greater than that expected by chance [10]. More powerful versions of Fitch test have appeared subsequently leading to a general framework for testing the molecular clock hypothesis for both DNA and protein sequences between two lineages, given an outgroup species [11].

More recently, new procedures for testing the molecular clock on multiple lineages simultaneously in a phylogenetic tree have been proposed. Some of these tests identify anomalous groups or lineages, whereas some merely test an entire tree for conformity to the hypothesis. The index of dispersion has been suggested as an estimator to test the molecular clock. The rationale is that when the number of substitutions follows a Poisson law the index of dispersion is equal to one [12]. The problems with this type of test have been extensively discussed [13,14] and since then the approach based on this estimator has been dismissed.

Despite the observation that the strict molecular clock hypothesis does not fully explain the substitution process, it still remains a promising concept and a powerful analytical tool in evolutionary biology. Therefore, testing the molecular clock in phylogenetic trees is an essential task. The problem is that if one assumes that the substitution process is Poisson then there is no homogeneity in the distribution of substitutions along a lineage, that is, even though the clock rate is constant, the variance is as large as the rate itself. In fact, in a very precise sense, a Poisson process is as heterogeneous as it is possible, meaning that it distributes dots "at random" over a half-infinite line and is often called the "completely random process" [15]. Takahata [16] observed that the rates of molecular evolution in several loci are more irregular than described by simple Poisson processes and therefore the clock is over-dispersed in these situations. Accordingly, statistical models for the over-dispersed molecular clock were proposed which suggested that the over-dispersion of molecular clock is due either to a major molecular reconfiguration led by a series of subliminal neutral changes or to selected substitutions fine-tuning a molecule after a major molecular change [17].

In this work we propose to directly verify Pauling-Zuckerkandl's assumption that the substitution process along the branches of a (strictly) clock-like phylogeny is approximately Poisson by testing the hypothesis that the distribution of the branch lengths a phylogenetic tree, when measured as the average number of substitutions, follows a Poisson law. Accordingly, we developed a procedure for testing the strict molecular clock in phylogenetic trees where the inference is made using a non-parametric goodness-of-fit test.

The method proposed here introduces two novelties: (1) it is based on an ensemble of trees, instead of using only one single consensus tree – this is quite natural from the Bayesian framework point of view. Indeed, the Bayesian inference procedure generates a posterior distribution over the set of phylogenetic trees, in the form of an ensemble of representative trees; (2) the use of well-known non-parametric goodness-of-fit test known as *Kolmogorov-Smirnov* (KS) *test* [18–20], modified to account for discrete variables in Poisson distributions with estimated parameter $\lambda$.

The classical KS test is performed given a sample of size $N$ of independent and identically distributed (IID) observations of the variable of interest one computes the *empirical cumulative distribution function* (ECD) of the observed data, defined as

$$F_N(x) = \frac{\text{number of elements in the sample which are} \leq x}{N} \quad (1)$$

where $x$ is a positive real number.

The *expected cumulative distribution* $F_E(x)$ is the cumulative distribution function corresponding to the expected probability distribution function of the variable of interest, and the test statistic is $D_{KS} = \sup|F_N(x) - F_E(x)|$, which is a measure of distance between the two distributions. The null hypothesis **H₀:** "$F_E(x) = F_N(x)$ for all $x$" is rejected if $D_{KS}$ exceeds a critical value $D_\alpha$ for a fixed significance level $\alpha$. In the classical KS test, the null distribution of $D_{KS}$ does not depend on the expected distribution $F_E(x)$ and is given by an explicit formula – tabulated critical values have been available from [21]. Moreover, the test has statistical power, or sensitivity, tending to 1 as the sample size tends to infinity [22], i.e., the probability of a Type II Error (false negative rate) goes to zero.

However, the universality of the null distribution of $D_{KS}$ comes at a price: (i) the test only applies to continuous distributions and (ii) the parameters of the expected distribution cannot be estimated from the data (the expected distribution must be completely specified in advance) – in fact, these two conditions are necessary to show that the distribution of $D_{KS}$ is independent of expected distribution. This seems to be a serious obstruction for the use of the KS test in practical applications. Indeed, the use of the tables associated with standard KS test when one of the conditions (i) or (ii) are not satisfied results in conservative *p*-values, in the sense that the probability of Type I Error (false positive rate) is smaller than that given by the standard table [23].

This difficulty has prevented the dissemination of the KS test in biology and other fields.

Nevertheless, it is possible to circumvent these limitations and modify the KS test to the case of discrete distributions with or without estimation of parameters [24–26]. The test statistic remains unchanged, but its null distribution is not universal anymore; unlike the continuous completely specified case, it depends on the type of the expected distribution – more specifically, it depends on the set of points of discontinuity of the expected distribution – and on the parameters that are estimated from the data. Nowadays, with the advent of fast and cheap computers, $p$-values and critical values for modified KS tests can be easily calculated.

In the particular case where the expected distribution is Poisson, the modified test is a procedure for the following null hypothesis **H₀:** "$F_N(x)$ is Poisson with estimated mean value $\lambda$", against the alternative hypothesis **H₁:** "$F_N(x)$ is not Poisson", where the Poisson parameter (mean value) $\lambda$ is estimated as the arithmetic mean of a list of non-negative integer numbers (the observed data).

The test is performed by calculating the KS statistics ($D$-score):

$$D_{\text{PKS}}(N) = \sup|F_N(x) - P(x,\lambda)| \quad (2)$$

where $P(x,\lambda)$ is the *cumulative distribution function* for the Poisson distribution with parameter $\lambda$, defined for all positive real values $x$ with $k \leq x < k+1$ for all integral values $k = 0,1,2,\ldots,\infty$, as

$$P(x,\lambda) = e^{-\lambda}(1 + \lambda + \lambda^2/2! + \ldots + \lambda^k/k!) \quad (3)$$

The null hypothesis is rejected if $D_{\text{PKS}}$ exceeds the critical value $D_\alpha$ for a fixed significance level $\alpha$. Now the critical value $D_\alpha$ must be obtained from the distribution of the statistic $D_{\text{PKS}}$ which depends on the fact that the empirical distribution function is expected to be Poisson with parameter $\lambda$. This procedure is usually referred as the *Poisson-Kolmogorov-Smirnov* (PKS) test.

Campbell and Oprian [27] computed tables for approximated critical values of the PKS test (see also [28]). One can eliminate the need for tables by performing the parametric bootstrap (Monte Carlo simulation) for the PKS test as described in [29,30]. However, neither the tables nor the parametric bootstrap for the PKS test provide the exact critical values and $p$-values. More recently, Frey [31] gives an algorithm for the computation of exact $p$-values and exact critical values for the PKS test.

The dual *coverage band*, also called *confidence bands*, for the PKS test can be constructed as follows. Given a fixed significance level $\alpha$ with corresponding critical value $D_\alpha$ define the functions: $U(x) = \min\{F_N(x) + D_\alpha, 1\}$ and $L(x) = \max\{F_N(x) - D_\alpha, 0\}$. Then, the pair $(L(x), U(x))$ is a $100(1-\alpha)\%$ *non-parametric coverage band* for $F_N(x)$. Thus, for every piecewise continuous function $H$ with jump discontinuities on the natural numbers, the probability that $L(x) \leq H(x) \leq U(x)$ for all $x$, conditional on $H$ having mean value $\lambda$, is greater than $(1-\alpha)$ [32,33].

Another variation of the KS test is the *two-sample Kolmogorov-Smirnov* (2KS) test. In this case, it is given two samples of independent and identically distributed (IID) real-valued observations of sizes $N$ and $M$, respectively, and the test examines whether the two samples come from the same continuous distribution or not. For each sample, one computes the respective ECDs $F_N(x)$ and $G_M(x)$ (defined by the right-handed side of equation (1) with $N$ replaced by $M$) and the test statistics ($D$-score) is defined as:

$$D(N,M) = \sup|F_N(x) - G_M(x)| \quad (4)$$

Here, the null hypothesis is **H₀:** "$F_N(x) = G_M(x)$ for all $x$" and the the alternative hypothesis **H₁:** "$F_N(x) \neq G_M(x)$ for some $x$". The null hypothesis is rejected if $D(N,M)$ exceeds a critical value $D_\alpha$ for a fixed significance level $\alpha$. As in the standard one-sample KS test, the null distribution of $D(N,M)$ does not depend on the cumulative distributions $F_N$ and $G_M$ when they are assumed to be continuous (that is, when the two samples are real-valued). Tabulated critical values have been available from Massey [34]. As before, it is possible to compute approximated critical values and $p$-values by parametric bootstrap and the coverage bands are given by the standard procedure [32,33].

In statistics, the term "non-parametric" has at least two different meanings. The first meaning refers to techniques that do not rely on data belonging to any particular distribution. In particular, it includes procedures that test hypotheses that are not statements about population parameters and procedures that make no assumption about the sampled population, also called *distribution free* procedures [35]. The second meaning refers to techniques that do not assume that the structure of a model is fixed. For instance, the simple procedures for testing the molecular clock hypothesis between two lineages of [10,11] are non-parametric in this second sense, given that they do not require information on the model of evolutionary change. The methods proposed in this paper are non-parametric tests in the first meaning described above.

## Results and Discussion

### The Poisson-Kolmogorov-Smirnov test for the strict molecular clock

Our first proposal for testing the (strict) molecular clock is to apply the Kolmogorov-Smirnov test for Poisson distributions with estimated parameter to an ensemble of phylogenetic trees that has been generated by the Bayesian inference method. The PKS test for the (strict) molecular clock phylogeny is performed as follows:
(1) Generate one ensemble of unconstrained phylogenetic trees. It is recommended to choose an "outgroup taxon" that should be used to root the tree. The "outgroup branch" must be removed before performing the test.
(2) Perform a burn-in discarding at least 25%. After the burn-in the log-likelihood scores stabilize, and therefore all trees are considered statistically equivalent.
(3) Extract all branch lengths of each tree and convert each real value into a non-negative integer, the average number of substitutions, by multiplying each branch length by the size of the alignment and rounding the value, producing an ensemble of discrete, integer-valued, branch lengths.
(4) Compute the ECD $F_N(x)$ using all the discrete branch lengths of the ensemble by formula (1). Here, the unadjusted sample size $N$ is a cut-off value that determine the number of trees used in the test (see below).
(5) Compute the discrete mean branch length $\lambda$ using the same set of branches from item (4), as the average value of all branch lengths and compute the expected cumulative distribution $P(x,\lambda)$ by formula (3).
(6) Compute the test statistics $D_{PKS}(N)$ by formula (2) with the unadjusted sample size $N$.
(7) Compute the adjusted PKS sample size $N_{ADJ}$ (see below).
(8) Compute the appropriate critical value $D_\alpha$ (for a fixed significance level $\alpha$) or the $p$-value, using the adjusted sample size $N_{ADJ}$. The simplest way to obtain PKS critical values or $p$-values is from the tables of Campbell and Oprian [27].
Conclusion: If $D_{PKS}(N) > D_\alpha$ or the $p$-value is smaller than the threshold (say $p < 0.01$) then the null hypothesis **is rejected**, that is, the phylogeny **is not clock-like**, more precisely, the clock-like phylogenetic model does not fit the data well enough (under-fitting). If $D_{PKS}(N) < D_\alpha$ or the $p$-value is bigger than the threshold (say $p > 0.01$) then the null hypothesis **is not rejected**, that is, the phylogeny **is clock-like**, more precisely, the clock-like phylogenetic can fit the data as well as the non-clock like phylogenetic model, and since the clock-like model has fewer parameters it is preferable.
Definition of PKS sample sizes: The unadjusted sample size is defined as $N = \tau B$, where $B$ is number of branches of the trees and $\tau$ is the least number of trees that satisfies the following two conditions: (i) $\tau \geq$ number of taxa; (ii) $D_{PKS}$ is minimal with respect to $\tau$, that is, the fit of the ECD is the best possible, given that condition (i) is satisfied. The adjusted sample size is defined as $N_{ADJ} = k N = k \tau B$, where the auto-correlation coefficient is defined as $k = T_{ESS}/T$, with $T$ the total number of trees generated by the MCMC sampler, after the burn-in and $T_{ESS}$ the effective sample size associated to the tree lengths (TL) computed by the MCMC sampler.

**The two-sample Kolmogorov-Smirnov test for molecular clock**

Our second proposal for testing the molecular clock is to apply the 2KS test with two ensembles of trees that have been generated by Bayesian phylogeny inference. The test is performed as follows:
(1) Generate one ensemble of unconstrained phylogenetic trees. It is recommended to choose an "outgroup taxon" that should be used to root the tree.
(2) Generate one ensemble of clock-constrained phylogenetic trees. If the clock being tested is the strict clock then the trees are rooted by default and hence the "outgroup taxon" should not be included.
(3) Perform a burn-in discarding at least 25%. After the burn-in, the log-likelihood scores stabilize, and therefore all trees in the ensemble are considered statistically equivalent.
(4) Extract all the branch lengths, as real numbers, of each tree in both ensembles.
(5) Compute the ECD $F_N(x)$ using all the branch lengths in the ensemble of unconstrained phylogenetic trees by formula (1). Here the 2KS sample size $N$ is the total number of branches used to compute the ECD $F_N(x)$ (see below).
(6) Compute the ECD $G_M(x)$ using all the branch lengths in the ensemble of clock-constrained phylogenetic trees by formula (1) with $N$ replaced by $M$. Here the 2KS sample size $M$ is the total number of branches used to compute the ECD $G_M(x)$ (see below).
(7) Compute the test statistics $D(N,M)$ using formula (4).
(8) Compute the adjusted sample sizes $N_{ADJ}$ and $M_{ADJ}$ (see below).
(9) Compute the appropriate critical value $D_\alpha$ (for a fixed significance level $\alpha$) or $p$-value, using the adjusted sample sizes $N_{ADJ}$ and $M_{ADJ}$. In the 2KS test, critical values $D_\alpha$ or $p$-values are the usual ones (see Massey [34]).
Conclusion: If $D(N,M) > D_\alpha$ or the $p$-value is smaller than some threshold (say $p < 0.01$) the null hypothesis **is rejected**, that is, the phylogeny is **not clock-like**, more precisely, the clock-like phylogenetic model does not fit the data well enough (under-fitting). If $D(N,M) < D_\alpha$ or the $p$-value is bigger than the threshold (say $p > 0.01$) then the null hypothesis **is not rejected**, that is, the phylogeny **is clock-like**, more precisely, the clock-like phylogenetic can fit the data as well as the non-clock like phylogenetic model.
Definition of 2KS sample sizes: The unadjusted sample sizes are defined as $N = M = \tau B$ and where $B$ is number of branches of the trees and $\tau$ is the least number of trees that satisfies the following two conditions: (i) $\tau \geq$ number of taxa; (ii) $D_{PKS}$ is minimal with respect to $\tau$, that is, the fit of the ECD is the best possible, given that condition (i) is satisfied. The adjusted sample sizes are defined as $N_{ADJ} = k_C N$ and $M_{ADJ} = k_U M$, where the auto-correlation coefficient $k_C$ is obtained from the clock-constrained ensemble and the auto-correlation coefficient $k_U$ is obtained from the unconstrained ensemble, associated to the tree lengths (TL) computed by the MCMC sampler, as before.

**Poisson mixtures**

When the tree topology is highly unbalanced, with some branches much longer than others, it is expected that the long branches accumulate more changes than the short branches. For example, in a tree ((A,B),C), if it is perfectly clock-like, branches "A" and "B" will be equal in length, but branch "C" will be longer than both. The interior branch is unlikely to be exactly the same length as the three exterior branches. In such cases, if the tree is clock-like, we would expect that the (discretized) branch length distribution is a *mixture* of several Poisson distributions.

In such cases, the PKS test is expected to reject the null hypothesis. This does not mean that the tree is not clock-like, even though the branch length distribution is multi-modal. It would be more appropriate to consider that the test was inconclusive and try other approaches: (i) directly test the null hypothesis of a Poisson mixture using the branch lengths of the consensus tree to estimate the several Poisson parameters; (ii) apply the PKS test to appropriate sub-trees; (iii) apply the 2KS test.

**Performing the tests**

We have performed the test proposed here with several ensembles of trees obtained from three types of sequence data: a set of simulated sequences, a set of sequences generated by *in vitro* evolution [36], three small datasets of real sequences and a large data-set of real sequences [37] of fungal 18S rRNA gene (18S rDNA).

**Simulated data**

We generated a dataset of simulated sequence containing 9 sequences with 10,000 nucleotides each, using the Kimura 2-Parameter (K2P) nucleotide substitution model [38] to evolve the sequences. Since the counting process associated to the K2P substitution model is a Poisson process [12] we expected the branch lengths to follow a Poisson law. The tree used to evolve the nucleotide sequences was the "nine-taxon tree" of [39] (**Fig 1**). The tree has a long branch corresponding to the "outgroup" {O} and a sub-tree, the "ingroup" {S1,...,S8}, consisting of 8 taxa. We generated an ensemble of 1,000 trees (obtained from 2,000 trees with 50% burn-in). The topology was fixed to be the nine-taxon tree, with the taxon "O" as the outgroup and the K2P model was used to estimate the substitution rates.

First, we observe that the mean branch length of the ingroup is $\lambda$=5.75, while the mean branch length of the outgroup is $\lambda$=25.18; the outgroup branch is approximately 5 times longer than the branches of the ingroup. Taking into account that the tree is unrooted, this is consistent with the fact the long branches accumulate an average of five times more substitutions than the branches of the ingroup. The mean branch length of the full tree is $\lambda$=7.01, is approximately equal to the weighted average of branch lengths of the ingroup and the outgroup: 14/15 x 5.75 + 1/15 x 25.18 = 7.03. This fact suggests, as expected, that the branch length distribution of the full tree is a (14/15,1/15) weighted mixture of two independent Poisson distributions with parameters $\lambda_1$=5.75 and $\lambda_2$=25.18, respectively. The hypothesis of Poisson mixtures is further strengthened by comparing the ECD obtained from the branch lengths of full tree ensemble with a Poisson cumulative distribution with parameter $\lambda$=7.01 and the cumulative distribution of a Poisson mixture with parameters $\lambda_1$=5.75 and $\lambda_2$=25.18 with weights (14/15, 1/15) (**Fig 2a-2b**).

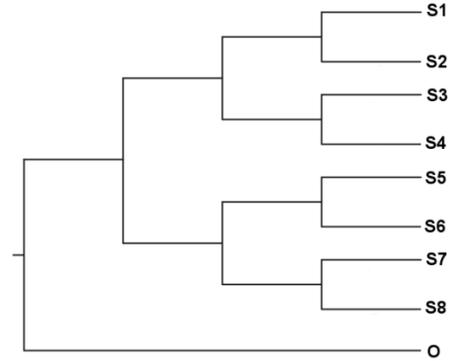

**Fig 1. The "nine-taxon tree" used to evolve the simulated nucleotide sequences.** Dataset of simulated sequences containing 9 sequences with 10,000 nucleotides each, using the Kimura 2-Parameter (K2P) nucleotide substitution model [38] to evolve the sequences.

We have performed the PKS test with the full tree, with the purpose of illustrating the situation when the observed data is a (bimodal) Poisson mixture. As expected the PKS test concludes that the phylogeny was not clock-like, with a *p*-value < 0.001%. On the other hand, when we removed the "outgroup branch" and performed the test with the ingroup the PKS test concluded that the phylogeny was clock-like, with *p*-value of 98% (**Table 1**). We also performed the test with the outgroup and the PKS test concluded that the tree is clock-like, with *p*-value of 34% (**Table 1**). The ECDs of branch lengths obtained from the ingroup and the outgroup ensembles with the respective expected cumulative Poisson distributions are shown in **Fig 2c-2d**.

**Table 1.** PKS test for strict clock on the simulated data (9 taxa) (1,000 trees, *k*=0.50).

|  | Full Tree | Ingroup | Outgroup |
| --- | --- | --- | --- |
| Mean Branch Length ($\lambda$) | 7.01 | 5.75 | 25.18 |
| $D_{PKS}$ | 0.15 | 0.01 | 0.06 |
| $\tau(B)$ | 33 (15) | 49 (14) | 182 (1) |
| $N_{ADJ}$ | 247 | 343 | 91 |
| Critical Value (1%) | 0.05 | 0.04 | 0.10 |
| *p*-value | < 0.00001 | 0.98 | 0.34 |
| Power estimate | > 0.99 | > 0.99 | > 0.99 |

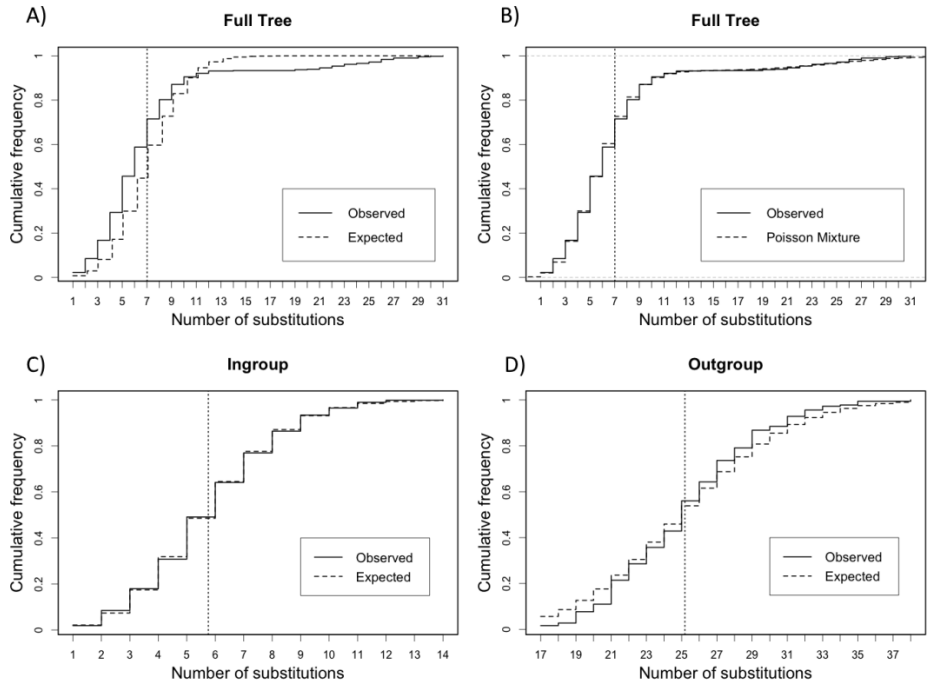

**Fig 2. Empirical cumulative distributions and expected cumulative distributions of the simulated sequences ensemble.** Vertical dotted line indicates the mean value. Panels (**a**) and (**b**) display the ECD of the full tree against the expected cumulative Poisson distribution with mean branch length $\lambda$=7.01 and the expected cumulative Poisson mixture, respectively. Panel (**c**) displays the ECD of the ingroup against the expected cumulative Poisson distribution with mean branch length $\lambda$=5.75. Panel (**d**) displays the ECD of the outgroup against the expected cumulative Poisson distribution with mean branch length $\lambda$=25.18.

### *In vitro* evolution data

The 16 sequences of 2,238 nucleotides were generated by four-step serial bifurcate PCR method, where the ancestor sequence 18S rRNA gene (18S rDNA) evolved *in vitro* for 280 nested PCR cycles [36]. The real phylogeny obtained in the experiment, with the number of substitutions on each branch is shown in **Fig 3**.

We generated an ensemble of 1,000 trees (obtained from 2,000 trees with 50% burn-in). The additional "outgroup taxon" that was removed during the extraction of branch lengths, and we used the General Time Reversible (GTR) model to estimate the substitution rates (actually, the best substitution model fitting the *in vitro* evolution obtained in [36] is not even time-reversible).

We have performed the PKS test with the ensemble of trees and with the real tree. The PKS test concluded that phylogeny was not clock-like, with a *p*-value < 0.001%. On the other hand, when performed just with the consensus tree the PKS test concluded that the phylogeny was clock-like, with a *p*-value of 65% (**Table 2**). The ECDs of branch lengths with the respective expected cumulative Poisson distributions are shown in **Fig 4a-4b**. The incongruence found between the results of the test on the real tree and the ensemble of trees is most likely because the adjusted PKS samples size ($N_{ADJ}$ = 465) obtained for the ensemble of trees, even though being small in relation to the ensemble, was capable of providing enough information and power to the test. Another possibility is that the tree topology was not well balanced and hence a test for a Poisson mixture could provide another conclusion.

### Small data-sets of real sequences

In order to illustrate the method with highly unbalanced tree topologies we have performed both tests with three data-sets of real sequences: (1) the *ENV* viral gene of immunodeficiency Lentiviruses (9 sequences of 3,013 nucleotides), (**Fig 5**); (2) the *COX1* gene of Primates (9 sequences of 1,569 nucleotides), (**Fig 6**); (3) the 18S rRNA gene (18S rDNA) of Ascomycete yeasts (17 sequences of 1,845 nucleotides), (**Fig 7**). For each data-set we generated 3 ensembles with 1,000 trees (2,000 trees with 50% burn-in) – one non-clock, one strict clock and one relaxed clock – each of them under the GTR model.

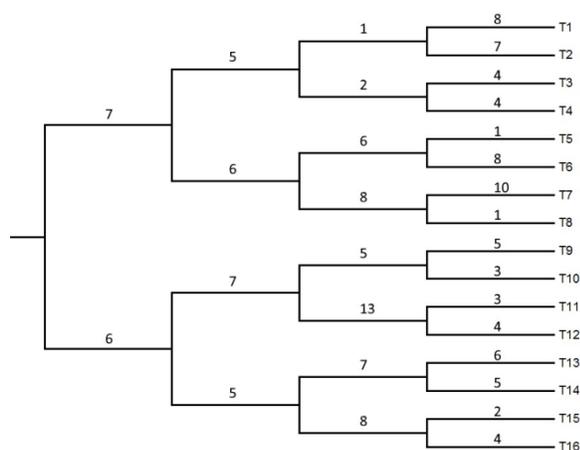

**Fig 3. The real phylogeny obtained by *in vitro* PCR evolution.** As described in Sanson et al. [36] with the number of observed substitutions along each branch.

We performed the PKS test with the three ensembles. The PKS test concluded that the phylogenies were not clock-like in all the three cases (**Table 3**). Since the trees are highly unbalanced this was expected as the branch length distributions are multi-modal in all cases. However, the PKS test could have concluded this because of the unbalanced topology and not because the tree is nor clock-like.

We performed the 2KS test for the unconstrained ensemble against the strict clock-constrained ensemble and the uncorrelated rates log-normal relaxed clock model [40]. The 2KS test concluded that the phylogenies were not strict clock-like in the three cases (**Table 4**). In all three cases, it is clear from the ECDs that the branch length distribution of the unconstrained ensemble is over-dispersed in relation to the strict clock-constrained ensemble. On the other hand, the 2KS test concluded that the phylogenies were clock-like for the relaxed clock model (**Table 5**). Nevertheless, it is apparent from the ECDs that for the Lentiviruses *ENV* the test was not overwhelmingly conclusive as in the other cases. The difference between the *ENV* and the other two cases can be seen by constructing the coverage bands for the three cases (we have removed the longer branches from the *ENV* ensemble). In fact, the 99% confidence band of the *ENV* ECD was very wide (0.125) compared with other two cases (0.062 for *COX1* and 0.091 for 18S rDNA), see **Fig 8**. The situation of *ENV* could be resolved by trying to fit other relaxed clock models and/or performing more powerful two-sample tests.

**Large data-set of real sequences**

Until now we have used some small data-sets to illustrate several aspects of the method, which might give the impression that the size of the data-set is a limitation of the method. This is not the case, the method works equally well, independently of the number of sequences.

**Table 2.** PKS test for strict clock on Sanson *et al.* [36] data (16 taxa) (1,000 trees, $k$ = 0.50).

|  | Ensemble of Trees (GTR) | Real Tree [36] |
|---|---|---|
| Mean Branch Length ($\lambda$) | 5.76 | 5.36 |
| $D_{PKS}$ | 0.07 | 0.07 |
| $\tau$ ($B$) | 31 (30) | 1 (30) |
| $N_{ADJ}$ | 465 | 30 |
| Critical Value (1%) | 0.04 | 0.16 |
| $p$-value | < 0.00001 | 0.65 |
| Power estimate | > 0.83 | > 0.83 |

We have performed the tests on the data-set of [37], consisting of an alignment of 134 sequences (with 1474 bp) of fungal 18S rDNA, of which 131 are representative of all groups of Fungi, 2 are representatives of animal (*Clathrina cerebrum*) and plants (*Sphagnum cuspidatum*) and 1 outgroup (*Developayella elegans*), see [37] for the complete list of taxa, accession numbers and the phylogeny. We generated 3 ensembles with 15,000 trees (20,000 trees with 25% burn-in) – one non-clock, one strict clock and one relaxed clock – each of them under the GTR model.

We performed the PKS test, which concluded that the phylogeny was not clock-like (**Table 6**). Since the tree is extremely unbalanced this was expected, as the branch length distributions are multi-modal in all cases. However, the PKS test could have concluded this because of the unbalanced topology and not because the tree is nor clock-like.

We performed the 2KS test for the unconstrained ensemble against the strict clock-constrained ensemble and the uncorrelated rates log-normal relaxed clock model [40]. The 2KS test concluded that the phylogeny was not strict clock-like. Moreover, the 2KS test concluded that the phylogeny was not clock-like for the relaxed clock model, as well (**Table 7**). From the ECDs it seems that the strict clock and the relaxed clock look very similar. We performed the 2KS test for the strict clock-constrained ensemble against the relaxed clock-constrained ensemble and, as expected, the test concluded that they are not the same (**Table 7**). It is clear from the ECDs that the branch length distribution of the unconstrained ensemble is highly over-dispersed in relation to the strict clock-constrained and the relaxed clock-constrained ensembles. In fact, the 99% confidence band of the unconstrained ensemble ECD excludes a large portion of the ECDs of the other two clock constrained ensembles (**Fig 9**). In this case, the conclusion indicates that other relaxed clock models should be considered.

**Comparison with other tests**

There are other statistical tests for the molecular clock, including relaxed clocks and they are of two types: (i) tests that are applied directly to a consensus tree or to the ensemble of trees generated by a standard run of the MCMC sampler (e.g., *Likelihood Ratio* (LR) test [41]); (ii) tests that require extensive additional computations beyond the standard run of the MCMC sampler (e.g., *Bayes Factor*). Since our test is of the first type we have performed a simple comparison between the PKS test and the LR test in order to illustrate their application.

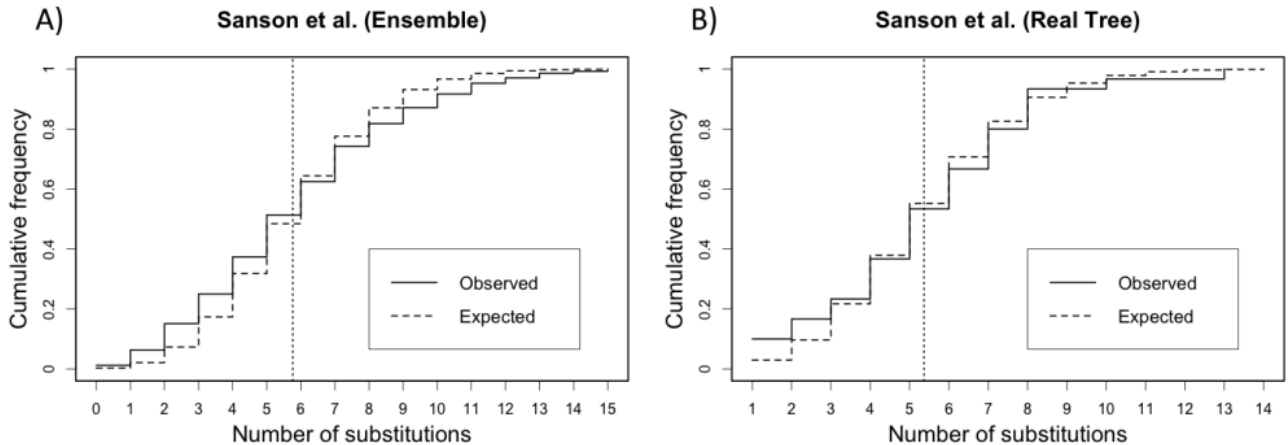

**Fig 4. Empirical cumulative distributions of the Sanson *et al.* [36] sequences ensemble.** Vertical dotted line indicates the mean value. Panels (**a**) and (**b**) display the empirical cumulative distribution of the observed data from ensemble of trees against the expected cumulative Poisson distribution with mean branch length *λ*=5.75 and the empirical cumulative distribution of the observed data from the real tree against the expected cumulative Poisson distribution with mean branch length *λ*=5.36, respectively.

Let us recall the main points of the Likelihood Ratio (LR) test [41]. Under the strict clock hypothesis (**H$_0$**), there are $S-1$ parameters corresponding to the ages of the $S-1$ internal nodes on a rooted tree with $S$ species. The more general non-clock hypothesis (**H$_1$**) allows every branch to have its own rate. Because time and rate are confounded, there are $2S-3$ free parameters, corresponding to the branch lengths in the unrooted tree. The clock model is equivalent to the non-clock model by applying $S-2$ equality constraints. If $L_0$ and $L_1$ are the log-likelihood values under clock and non-clock models, respectively, then $2\Delta L = -2(L_1 - L_0)$ is compared with the critical value from the $\chi^2$ (chi-square) distribution with $v = S-2$ degrees of freedom to decide whether the clock hypothesis is rejected (the chi-square distribution is the asymptotic null distribution of the test when $v$ is sufficiently large). In order to perform the LR test with trees generated by Bayesian inference one must compute two consensus trees (one for each ensemble).

We have performed the LR test with both the simulated data and *in vitro* evolution data in order to compare with the PKS test (**Table 8**). To perform the LR test we also needed to compute a clock-constrained consensus tree. In both cases, the null hypothesis that the phylogeny is strictly clock-like was not rejected. This is incongruent with the result of the PKS test performed on the ensemble of trees obtained for the *in vitro* evolution data, most likely because the effective samples size achieved in the PKS test was capable of providing enough information and power to the test.

It should be noted that the LR test does not examine whether the rate is constant over time. In fact, what is tested is the weather the hypothesis that all tips of the tree are equidistant from the root, with distances measured by the number of substitutions. Therefore, if the evolutionary rate has been equally accelerating (or decelerating) over time in all lineages, in the absence of a calibration the tree will be ultra-metric, although the rate is not constant. Second, the test cannot distinguish a constant rate from an average variable rate within a lineage, although the latter may be a more sensible explanation than the former when the clock is rejected and the rate is variable across lineages.

**Table 3.** PKS test for strict clock on three data-sets of real sequences *ENV, COX1* and 18S rDNA (1,000 trees, *k* = 0.50). BL = Branch Length, CV = critical value.

|  | Lentiviruses *ENV* | Primates *COX1* | Yeasts 18S rDNA |
|---|---|---|---|
| Mean BL (*λ*) | 752.3 | 186.2 | 26.5 |
| $D_{PKS}$ | 0.53 | 0.67 | 0.63 |
| *τ* (*B*) | 224 (13) | 17 (15) | 89 (31) |
| $N_{ADJ}$ | 1,458 | 127 | 1,382 |
| CV (1%) | 0.02 | 0.09 | 0.02 |
| *p*-value | < 0.00001 | < 0.00001 | < 0.00001 |

## Conclusion

In the present paper we introduce two non-parametric goodness-of-fit GOF tests based on the empirical cumulative distribution (ECD) to the context of testing for the molecular clock in phylogeny, by using branch lengths extracted from an ensemble of statistically equivalent trees. The use of an ensemble of statistically equivalent trees to compute the ECD from the observed data, instead of a consensus tree, relieves trees must be used to compute $F_N(x)$ – in our case this definition is not trivial, since it is not the original data (the sequence alignment) that is used in the test, but the trees generated by a MCMC sampler from it.

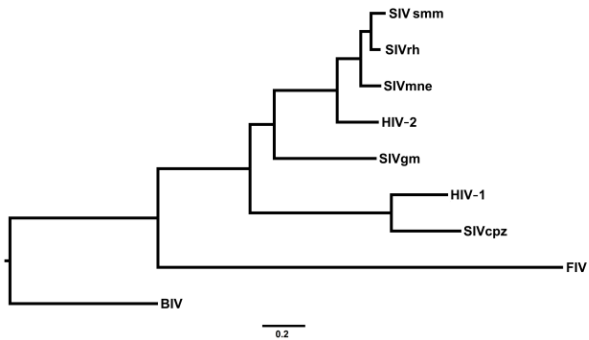

**Fig 5. The *ENV* phylogeny of immunodeficiency Lentiviruses.** Taxa and accession numbers are: HIV1 (K03455.1), HIV2 (M30502.1), BIV (M32690.1), FIV (M25381.1), SIVgm (U58991.1), SIVcpz (X52154.1), SIVmne (M32741.1), SIVrh (FJ842859) and SIVsmm (X14307.1). The scale bar indicates substitutions per position.

the effects of small sample size and issues related to lack of power and lack of information.

The PKS test allows to directly verify whether a phylogeny is clock-like following a Poisson law. The PKS test is very simple to apply and could even be performed simultaneously with the generation of the ensemble of trees, but it is more limited since it has a very restricted null hypothesis. The test may be extended to the case of Poisson mixtures in order to allow for more flexible null hypothesis.

The 2KS test is more flexible allowing for the investigation whether a phylogeny follows a relaxed clock model, but has more intricate usage since it requires the generation and comparison of two ensembles of trees. The 2KS test can distinguish the strict clock model from a relaxed clock model but it seems that it is not powerful enough to distinguish one relaxed clock model from another. Generally speaking, in both methods it is possible to replace the KS tests by another ECD based goodness-of-fit test, such as the Cramér-von Mises test or the Anderson-Darling test [42–44]. Another possible extension of these methods is to employ Bayesian non-parametric goodness-of-fit tests.

## Materials and Methods

### Theoretical Framework

In this section we provide the theoretical basis for the tests described in Section 2. Let us start with the computation of the ECD using the branch lengths obtained from an ensemble of trees. There are two points that must be clarified: (i) the fact that the trees used in the computation of $F_N(x)$ are *not* independent – this is an important issue, since one of the assumptions of all KS tests is that the data used to compute $F_N(x)$ is that it is IID and of failure this assumption may cause a pseudo-replication effect; (ii) what is the appropriate sample size $N$ for the KS test, that is, how many

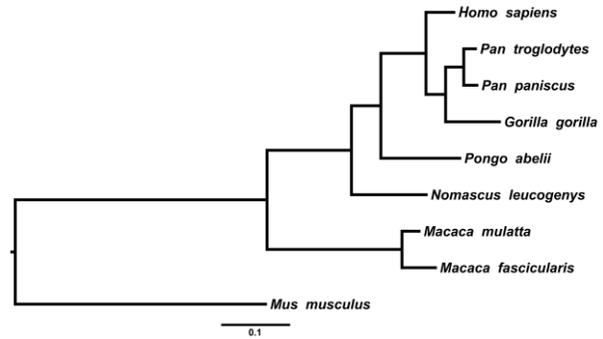

**Fig 6. The *COX1* phylogeny of Primates.** Taxa and accession numbers are: *Homo sapiens* (YP003024028.1), *Pan troglodytes* (NP008188.1), *Pan paniscus* (NP008201.1), *Gorilla gorilla* (YP002120661.1), *Pongo abelii* (YP007837.1), *Nomascus leucogenys* (YP008379101.1), *Macaca fascicularis* (YP002884228.1), *Mus musculus* (NP904330.1). The scale bar indicates substitutions per position.

The issue of non-independence of the trees is due to the fact that Bayesian phylogeny inference employs a Markov Chain Monte Carlo (MCMC) sampler to compute the trees. In addition, it is well known that critical values of the KS statistics are sensitive to sample auto-correlation, that is, the actual false positive rate tends to be higher when there exists a positive sample auto-correlation. As a result, direct use of the KS test might give misleading results when used with an auto-correlated sample, even though, it is possible to modify the test to account for sample auto-correlations [45,46]. Nevertheless, when sample auto-correlation is due to a process with short range memory (e.g. markovian process) there are two simple adjustments that allows for the application of KS test for IID samples to the case of auto-correlated samples [47]. Fortunately, both adjustments are already implemented in most of the MCMC samplers for Bayesian phylogeny inference. The first adjustment is called *thinning of sample*, which consists in discarding a fixed number of consecutive sampled trees after one tree is included in the ensemble so that the remaining trees are almost independent of each other. The second procedure is called *effective sample size* (ESS). When a sample has auto-correlations, the information contained in the data is (usually) less than the information contained in an IID sample with the same size. In other words, the number of equivalent independent observations is fewer than the actual sample size. The *effective sample size* (ESS) is the size of a putative IID sample that carries the same amount of information as the correlated sample.

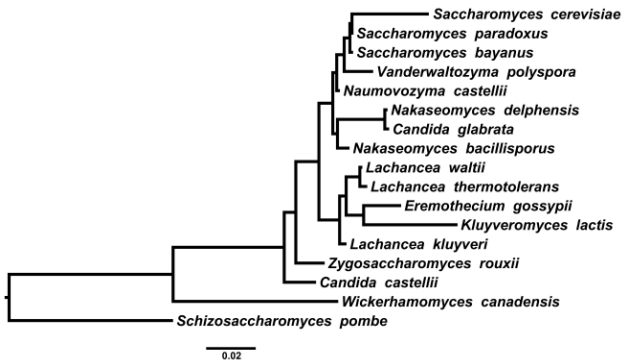

**Fig 7. The 18S rDNA phylogeny of Ascomycetes.** Taxa and accession numbers are: *Schizosaccharomyces pombe* (CU329672.1), *Lachancea waltii* (X89527), *Wickerhamomyces canadensis* (AB054565.1), *Eremothecium gossypii* (AY046265.1), *Zygosaccharomyces rouxii* (AY227011.1), *Kluyveromyces lactis* (HM009311.1), *Vanderwaltozyma polyspora* (JQ698890.1), *Nakaseomyces bacillisporus* (AY046252.1), *Nakaseomyces delphensis* (AY198400.1), *Candida glabrata* (KT229542.1), *Naumovozyma castellii* (HE576754), *Candida castellii* (AY046253.1), *Saccharomyces bayanus* (AY046227), *Saccharomyces cerevisiae* (JQ409454.1), *Saccharomyces paradoxus* (BR000309.1), *Lachancea kluyveri* (Z75580.1), *Lachancea thermotolerants* (CU928180.1). The scale bar indicates substitutions per position.

**Table 4.** Two-sample KS test for strict clock on three data-sets (*ENV* and *COX1* with 9 taxa and 18S rDNA with 17 taxa) of real sequences (1,000 trees, $k_U = 0.192$, $k_C = 0.116$).

|  | Lentiviruses *ENV* | Primates *COX1* | Yeasts 18S rDNA |
|---|---|---|---|
| $\tau(B)$ | 349 (15) | 940 (15) | 211 (31) |
| $N_{ADJ}$ (Non-clock) | 670 | 2,707 | 1,255 |
| $M_{ADJ}$ (Strict clock) | 404 | 1,635 | 758 |
| $D(M,N)$ | 0.21 | 0.11 | 0.11 |
| Critical Value (1%) | 0.10 | 0.05 | 0.07 |
| *p*-value | < 0.00001 | < 0.00001 | < 0.00001 |

In [48] the authors discuss the notion of effective number of independent observations in an auto-correlated time series, specifically with respect to estimating the mean and the variance of a distribution. Furthermore, it has been demonstrated that several hypothesis tests can be modified for auto-correlated data using an ESS adjustment. For example, [49] proposed a modified Mann-Kendall trend detection test with ESS adjustment. Similar to the KS test, the original Mann-Kendall test is a non-parametric test based on the IID sample assumption, which is not satisfied by most of time series data. The modification proposed in [49] is simply to replace the actual sample size by the effective sample size in the computation of the *p*-value or the critical value. With the the ESS adjustment the modified Mann-Kendall provides the correct rejection rate. Here we propose the same type of ESS adjustment for the KS test, by replacing the actual sample size by the effective sample size computed by the MCMC sampler for the tree length (TL) estimation. It should be remarked that in a strict clock-constrained tree the branch lengths are not independent, as well. In fact, the branch lengths are constrained in such a way that the distance from the root to the tips is the same for all tips. In this case, the ESS adjustment in the 2KS takes into account the lack of independence of the branch lengths by employing the adjustment with the coefficient $k_C$ associated to the tree length (TL) estimation.

Regarding the definition of appropriate sample size of the KS test, the reason why this is an issue here is because the the ensemble of trees used as the sample for the tests is generated by a MCMC sampler from the original data, and so the sample size *N* may, in principle, be arbitrarily large. But this would be an artifact, since the input for the computation of the ensemble is a finite set of finite sequences and one cannot extract an arbitrary amount of information form it by computing an arbitrarily large number of sample trees. Moreover, it is known that the critical value $D_\alpha = D_\alpha(\lambda, N)$ of the PKS test, as a function of the sample size, goes to zero as fast as $N^{-1/2}$, when *N* goes to infinity [50,51]. On the other hand, since $D_{PKS}(N)$ does *not* go to zero, when *N* goes to infinity (because the original data contains a finite amount of information), it is possible to increase the number of trees used to compute $D_{PKS}$ in such a way that the null hypothesis is always rejected, thus, rendering the test completely useless. In order to circumvent this difficulty we proposed the definition of the sample size in terms of the least number of trees *τ* such that $D_{PKS}$ is minimal with respect to *τ*, given that *τ* ≥ the number of taxa (this last condition ensures that the ensemble of trees is not too small). Then, if the null hypothesis is rejected under these conditions, increasing the sample size will not change the result, although it would artificially inflate the power of the test. In this case, the use of the ESS adjusted sample size $N_{ADJ}$ ensures that the power is not inflated. On the other hand, if the null hypothesis is not rejected, increasing the sample size would artificially revert this result. Hence, in this case, the use of $N_{ADJ}$ ensures that this reversion does not occur. Failure to reject the null hypothesis have two different meanings, depending on the magnitude of $N_{ADJ}$: (i) if $N_{ADJ}$ is sufficiently large and (**H₀**) is not rejected then it is possible that the null hypothesis is actually true; (ii) if $N_{ADJ}$ is small and (**H₀**) is not rejected then the test is inconclusive due to lack of information in the data from which the ensemble of trees was generated.

It is convenient to be able to estimate the power of the KS test in order to evaluate if the test correctly rejects the null hypothesis when the alternative hypothesis is true, equivalently, the probability of accepting the alternative hypothesis when it is true. However, it is very difficult to compute the asymptotic distribution of the KS statistics under the alternative hypothesis and thus it is difficult to compute the power of the KS test. It is possible, nevertheless, to find a lower bound for the power and use it to evaluate the asymptotic power of the KS test [22,52]. Although, this lower bound is conservative for discrete versions of the KS test [53], which includes the PKS version used here, it still may be used to evaluate the power of the test.

**Table 5.** Two-sample KS test for relaxed clock on three data-sets (*ENV* and *COX1* with 9 taxa and 18S rDNA with 17 taxa) of real sequences (1,000 trees, $k_U = 0.192$, $k_C = 0.099$).

|  | Lentiviruses *ENV* | Primates *COX1* | Yeasts 18S rDNA |
|---|---|---|---|
| $\tau(B)$ | 349 (15) | 940 (15) | 211 (31) |
| $N_{ADJ}$ (Non-clock) | 670 | 2,707 | 1,255 |
| $M_{ADJ}$ (Relaxed clock) | 354 | 1,395 | 647 |
| $D(M,N)$ | 0.043 | 0.005 | 0.007 |
| Critical Value (1%) | 0.107 | 0.050 | 0.070 |
| $p$-value | 0.773 | 0.999 | 0.999 |

Historically, KS tests have only been used as goodness-of-fit tests for continuous distributions, while the *chi-square test* has been commonly employed for discrete data. In [54] the author gives a comprehensive review of both and their competitors. The chi-square test statistics may also be written as a measure of discrepancy between the ECD $F_N(x)$ and expected cumulative distribution $F_E(x)$. However, it does not take into account the natural ordering among the observations, a fact exploited in analysis of attribute data. More specifically, the chi-square test statistics is invariant under permutations. In contrast, the KS test statistics is sensitive to the over-weighting or under-weighting of any tail or segment of the empirical distribution relative to the hypothesized distribution. It is for this reason that KS tests derive their greater advantages and is the main motivation behind several efforts to adapt KS tests to discrete data.

**Table 6.** PKS test for strict clock on Padovan *et al.* [37] data (134 taxa) (15,000 trees, $k$ =0.04).

|  | Non-clock ensemble |
|---|---|
| Mean Branch Length ($\lambda$) | 15.0 |
| $D_{PKS}$ | 0.47 |
| $\tau(B)$ | 377 (264) |
| $N_{ADJ}$ | 4,366 |
| Critical Value (1%) | 0.01 |
| $p$-value | < 0.00001 |

Finally, it is interesting to observe that the idea behind the likelihood ratio (LR) test of examining if all tips of the tree are equidistant has an analogue in the non-parametric setting proposed here. For each tree in the ensemble, a tree branch randomly picked and the distance from the root computed. Then, the ECD of distances from the tips to the root is computed and tested if the distribution follows a Poisson law (the same discussion about sample size and ESS adjustment should apply here). A hint that this procedure might work is provided by the analysis involving the outgroup ECD in the simulated example, which is the distance from the root to the tip and is very close to a Poisson distribution. The potential importance of this variation in the PKS test, is beyond the scope of the present study and may be investigated in a future work.

**Table 7.** Two-sample KS tests for strict and relaxed clock on Padovan *et al.* [37] data (134 taxa) (15,000 trees, $k_U$ =0.04, $k_C$ =0.19 (strict clock), $k_C$ =0.05 (relaxed clock)).

|  | Non-clock x Strict | Non-clock x Relaxed | Strict x Relaxed |
|---|---|---|---|
| $\tau(B)$ | 274 (267) | 274 (267) | 274 (267) |
| $N_{ADJ}$ | 2,926 | 2,926 | 13,900 |
| $M_{ADJ}$ | 13,900 | 3,657 | 3,657 |
| $D(N,M)$ | 0.078 | 0.059 | 0.044 |
| Critical Value (1%) | 0.033 | 0.040 | 0.030 |
| $p$-value | < 0.000001 | 0.000019 | 0.000020 |

**Table 8.** Likelihood ratio tests (1% significance level).

|  | Clock ($L_0$) | Non-clock ($L_1$) | $2\Delta L=-2(L_1-L_0)$ | $\chi^2 (S-2)$ |
|---|---|---|---|---|
| Simulated | −1,476.80 | −1,477.02 | 0.44 | 18.47 (8) |
| Sanson *et al.* | −4,312.59 | −4,321.71 | 18.24 | 29.14 (14) |

### Software and computation resources

The alignments were made with **Clustal W2** [55], the phylogenies were computed with **MrBayes 3.2.6** [56] and the model selection was computed with **jModelTest 2** [57]. The statistical software **R 3.3.2** [58] was used to prepare figures of ECDs and confidence bands, compute critical values and *p*-values. Phylogenetic tree manipulation and branch length extraction was made with the **R** package **APE 3.5** [59]. The simulation of nucleotide sequence was made with **Seq-Gen 1.3.2** [60]. Alternatively, we have created a simple program to perform the one-sample PKS test. The program imports the output files generated by **MrBayes**, extract the branch lengths from the ensemble of trees. The PKS critical values are computed using the tables of Campbell and Oprian [27]. The user may choose one of the three possible significance levels $\alpha$: 10%, 5% and 1%. The program outputs the following information: (i) the mean value and variance of the log-likelihood scores of the ensemble of trees; (ii) the mean and the variance of the ECD of branch lengths, measured in number of substitutions; (iii) the PKS statistics $D_{PKS}$ and the PKS critical value $D_\alpha$ for the chosen significance level. Finally, it is possible to plot the ECD of branch lengths and the expected Poisson cumulative distribution. The implementation of the PKS test is done in **Python 2.7** [61] with the libraries **Numpy**, **Pylab**, **Matplotlib**, **Tkinter**.

### Program availability

The source code of program PKS, that implements the method here described, is available at GitHub (github.com/FernandoMarcon/PKS_Test).

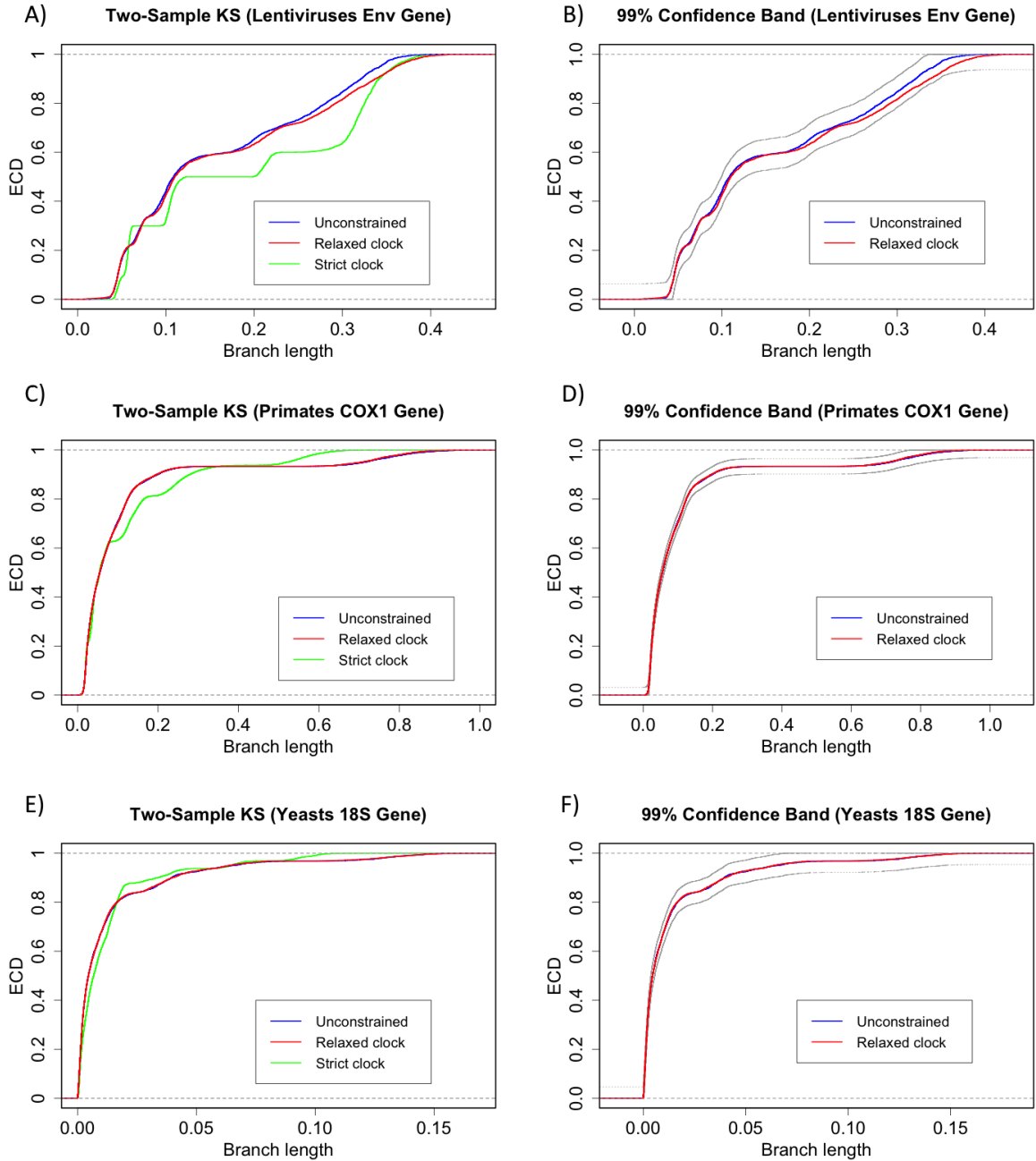

**Fig 8. Empirical cumulative distributions and 99% confidence bands for the three data-sets of real sequences.** The confidence band (the two black curves) is constructed around the ECD of the unconstrained ensemble. In the *COX1* and 18S rDNA, the blue curve is underneath the red curve.

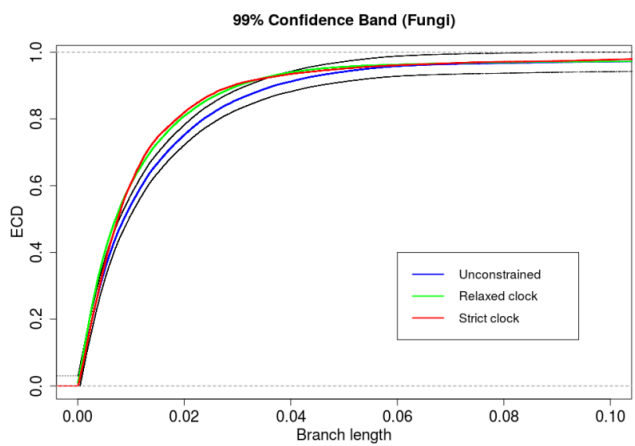

**Fig 9. Empirical cumulative distributions for the unconstrained (blue), strict clock (red) and relaxed clock (green) and a 99% confidence band for the large data-set of real sequences.** The confidence band (the two black curves) is constructed around the ECD of the unconstrained ensemble.

## Data availability

All alignments and MrBayes command blocks used in the analyses described are available as Supporting Information as Compressed/ZIP file Archive.
https://www.dropbox.com/s/n5h5ksl9owzu7g8/pks-alignments.zip?dl=0

## Acknowledgements

The authors thank Dr. Joseph Felsenstein for critical review of the manuscript and valuable suggestions regarding the sets of branch lengths as IID variables and effects of covariation and long branches. This work was supported by grants from Brazilian government agencies FAPESP (2013/07838-0) and CNPq to MRSB.